\documentclass[fleqn,10pt]{wlscirep}
\usepackage[utf8]{inputenc}
\usepackage[T1]{fontenc}
\usepackage{pdfpages}

\makeatletter
\AtBeginDocument{\let\LS@rot\@undefined}
\makeatother
\title{Edge modes in 1D microwave photonic crystal}

\author[1]{Aleksey Girich}
\author[1,2*]{Liubov Ivzhenko}
\author[1]{Ganna Kharchenko}
\author[1]{Sergey Polevoy}
\author[1,3,4]{Sergey~Tarapov}
\author[2]{Maciej Krawczyk}
\author[2]{Jarosław W. Kłos}
\affil[1]{O. Ya. Usikov Institute for Radiophysics and Electronics NAS of Ukraine, Kharkiv,Ukraine}
\affil[2]{ ISQI, Faculty of Physics, Adam Mickiewicz University Poznań, Poland}
\affil[3]{Gebze Technical University, Gebze, Kocaeli,  Turkey}
\affil[4]{ V.N. Karazin Kharkiv National University, Kharkiv,  Ukraine}

\affil[*]{ivzhenko@amu.edu.pl}

\begin{abstract}
The microstrip of modulated width is a realization of a one-dimensional photonic crystal operating in the microwave regime. Like any photonic crystal, the periodic microstrip is characterised by the presence of frequency bands and band gaps that enable and prohibit wave propagation, respectively. 
The frequency bands for microstrip of symmetric unit cell can be distinguished by $0$ or $\pi$ Zak phase. 
The sum of these topological parameters for all bands below a given frequency gap determines the value of the surface impedance and whether or not edge modes are present at the end of the microstrip. We demonstrate that edge modes are absent in a finite microstrip terminated at both ends in the centres of unit cells, but they can be induced by adding the defected cells. Edge modes present at both ends of the microstrip enable microwave tunneling with high transitivity in the frequency gap with or without a change in phase. This has been demonstrated experimentally and developed in detail using numerical simulations and model calculations.  The investigated system, with a doublet of edge modes in the frequency gap, can be considered as a narrow passband filter of high selectivity.
\end{abstract}
\begin{document}

\flushbottom
\maketitle

\thispagestyle{empty}


\section*{Introduction}

The structures with periodic modulation of geometrical or material parameters in one dimension can be treated as artificial one-dimensional (1D) crystals. 
For such a systems, one can address the fundamental question linking the topology of the band structure and the condition under which edge states can be observed. 
The topological properties of the bands, such as a Zak phase\cite{Zak82,zak89},
are used to indicate in which frequency gaps the edge states can be observed\cite{atala_direct_2013, Xiao14}. 
The experimental and theoretical studies of edge states in 1D artificial crystals have been carried out for photonic\cite{Yeh_77, Vinogradov06,Klos07,Malkova09,Wang16,Yilmaz18,Wang_2019, Henriques}, phononic\cite{xiao_geometric_2015,yin_band_2018,Li}, plasmonic\cite{wang_zak_2018,poli_selective_2015}, magnonic\cite{Rychly17} and electronic systems\cite{Kucharczyk,Klos03}.
The microwave systems\cite{marpaung_integrated_2019} in the form of periodic microstrips\cite{Kee99,Tarapov12,Chernovtsev_2007,Zhu,wu_kind_2023}, usually studied by electrical means, also attract the attention of the researchers interested in the study of the bulk-edge correspondence and the potential applications of edge states\cite{poli_selective_2015, Nakata2020}.  

The bulk parameters of a periodic microstrip consisting of two alternating sections of different widths determine the dispersion relation of electromagnetic waves. For truncated structures with given bulk properties, we can define the so-called surface impedance\cite{Xiao14,Nakata2020}, which can be used to determine the conditions for the existence of edge (or interface) states at the microstrip ends (or at the junction of two microstrips). The surface impedance is strictly related to the Zak phase, and thus can be used to determine the necessary condition for the existence of the edge (interface) state\cite{Nakata2020,mieszczak2022}. These recent studies have focused on the interface states between two periodic microstrips, their topological properties and their potential applications. However, the edge microwave states localised at the edges of the periodic microstrip connected directly to input (or output) ports remain unexplored in terms of their topological properties and applications.

In this work we study experimentally and numerically the edge states in finite microstrip of modulated width -- see Fig.~\ref{fig:system}. We show that the microwave edge states cannot exist at the ends of a periodic microstrip composed of centrosymmetric unit cells and terminating in symmetry points. We prove that this property is the result of the topological property of the band structure (Fig.~\ref{fig:dispersion}), i.e. the Zak phases of successive bands, and its relation to the surface impedance at the symmetry point. To induce the edge states, it is necessary to modify the shape of the first and last cells of the microstrip. When such a modification is introduced, the surface impedance in the gap can be matched to the impedance of the feed ports, and the narrow transmission band at frequencies within the band gap can be measured (Fig.~\ref{fig:band_structure}). Importantly, the edge states form symmetric and antisymmetric pairs, allowing microwave signals to be tunneled through the finite microstrip with 0 or $\pi$ phase shift (Fig.~\ref{fig:profiles}).

The paper is structured as follows. In the next section, we present the structure under investigation and the methods used in our study. We then analyze the band structure as a function of the bulk parameters -- in these studies, we determine the symmetry of the Bloch functions at the edges of the bands and the Zak phases for successive bands. Next, we show that the edge modes cannot exist for the microstrip terminated at the symmetry points and investigate how the modifications of the first and last cells of the microstrip induce the edge states and tune their frequencies. The studies are complemented by the analysis of the corresponding lumped element model. At the end of the paper, the short section Discussion summarizes the main results of the paper. Finally, the experimental and numerical methods are described.

\section*{System}

\begin{figure}[ht]
\centering
\includegraphics[width=\linewidth]{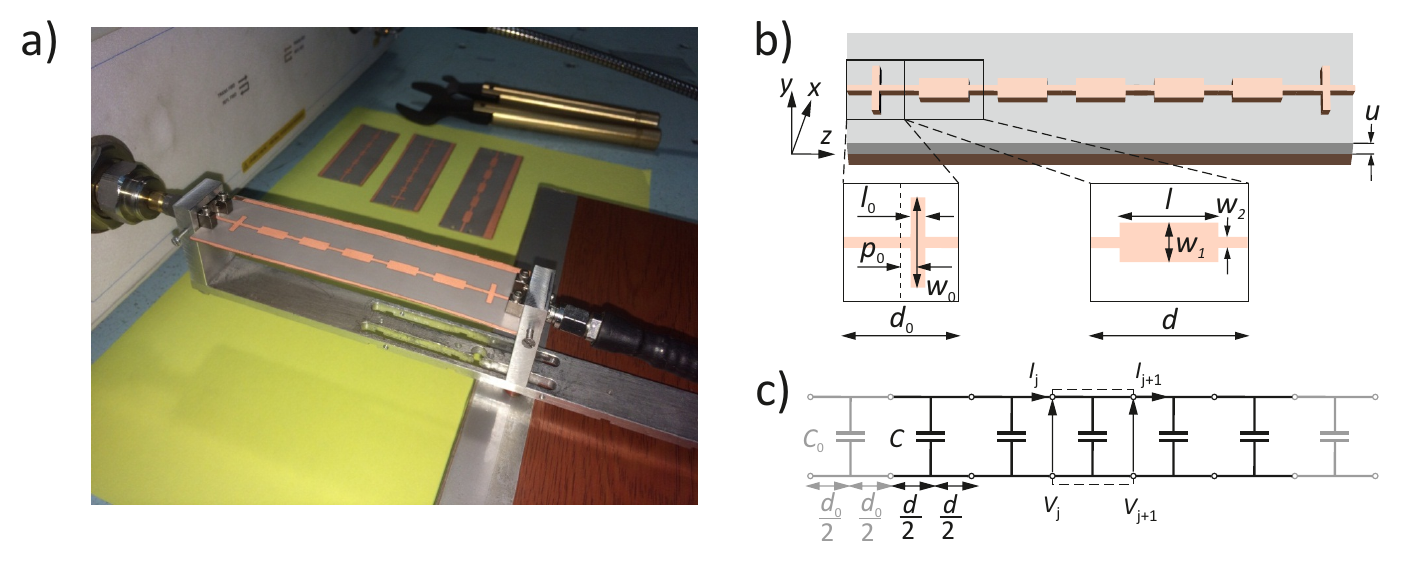}
\caption{Periodic microstrip under investigation. (a) The photo of the experimental setup showing the microstrip connected to a vector network analyzer with standard $Z_0 =50\; \Omega$ ports to measure the $S_{21}$ spectrum.
b) The geometry of the system used in the numerical simulations. The microstrip, composed of five bulk cells and two edge cells, is placed on the substrate of thickness $u$ with metallized bottom surface, which plays the role of the ground plane. 
The bulk and edge cells differ in size: $d$, and $d_0$, and in geometric parameters: ($l$,$w$), and ($l_0$, $w_0$), respectively.
For the edge cell, we also considered the shift $p_0$ of the wider segment with respect to the center of the cell.
(c) The lumped element model of the finite periodic microstrip, where the bulk and edge cells are represented by the symmetric two-port networks. Each network is composed of two phase shifters (resulting from the acquisition of the phase at the distance $d/2$ and $d_0/2$, respectively) and shunt susceptances (resulting from the presence of the capacitance $C$ and $C_0$, respectively, in each cell of the system).}
\label{fig:system}
\end{figure}

The experimentally investigated system is a microwave transmission line in the form of a periodically width-modulated microstrip composed of seven unit cells, as shown in Fig.~\ref{fig:system}(a). Each of (five) bulk cells (Fig.~\ref{fig:system}(b)) consists of the section of wider microstrip (width $w_1$ = 4 mm) of length $l$ = 10~mm connected to the neighboring cells by narrower sections (width $w_2$ = 1.12 mm). The length of the bulk unit cell $d$ is 16 mm. The system ends with (two) edge cells, which have different dimensions from the bulk cell (width $w_0$, length $l_0$ of the wider segment, the total length $d_0$, and the shift $p_0$ of the wider segment with respect to the center of the cell). The values $w_0$, $l_0$, $d_0$ and $p_0$ are set to 9 mm, 1.5 mm, 11.5 mm and 0 mm, respectively. For reference, we also studied the system without edge cells of modified sizes, i.e. composed of five cells of the same dimensions. The microstrip is deposited on the insulating substrate characterized by a dielectric constant of $\varepsilon = 2.2+\text{i}\,0.0012$ and a thickness of $u=0.381$~mm. The bottom side of the substrate is metallized and serves as the ground plane. The microstrip and the ground plane are made of copper with a conductivity of $\sigma=58$ kS/m and a thickness of 35 $\mu$m. 

A vector network analyser was used to measure the transmission spectra of the microwaves in the frequency range 8 - 20 GHz (see the photo of the experimental setup in Fig.~\ref{fig:system}(a)). The input and output of the microstrip are connected to a vector network analyzer with standard $Z_0=50$ $\Omega$ impedance ports to measure the $S_{21}$ transmission spectrum.
For the theoretical studies, full electromagnetic simulations were carried out using {\em CST Microwave Studio} with the same parameters as in the experimental studies. The experimental and numerical results were then complemented by semi-analytical calculations based on the lumped element model Fig.~\ref{fig:system}(c). More details on the methodology used can be found in the Methods section.

\section*{Results}
\subsection*{Measurements}

The transmission spectrum for the system, which consists of five identical unit cells, is presented with the red line in Fig.~\ref{fig:band_structure}(c), while the transmission spectrum for the microstrip with edge cells of modified sizes is shown in Fig.~\ref{fig:band_structure}(d) also with the red line. The results presents the transmission $|S_{21}(f)|$ in the range of 8-18 GHz. For the microstrip with undisturbed ends, a significant decline of the $|S_{21}|$ parameter is noted up to -30~dB within the 11.5-15.0~GHz frequency range (Fig.~\ref{fig:band_structure}(c). The spectrum for the microstrip with modified edge cells is significantly different. In the range 11.5-15.0~GHz, there is a peak at 13.5~GHz of relatively high $|S_{21}| = -10$~dB,  transmission. Additionally, the transmission for the frequencies larger than 15~GHz is significantly attenuated, and the oscillatory character of $|S_{21}|$ is observed at frequencies both higher than 11.5~GHz and  lower than 15~GHz. 

Clearly, these spectral changes result from the addition of the edge cells of modified sizes. In the upcoming sections, we will demonstrate with numerical simulations and model calculations that the suppressed transmission between 11.5 and 15.0~GHz is due to the band gap introduced by the microstrip's periodicity, whereas the  transmission peak about 13.5~GHz results from the microstrip's edge modes. In addition, we will argue that these edge mods result from the topological characteristics of the bands in an unmodified periodic microstrip, but necessarily supplemented with modified edge cells.

\subsection*{Band structure and topological properties of infinite system}

\begin{figure}[!ht]
\centering
\includegraphics[width=0.9\linewidth]{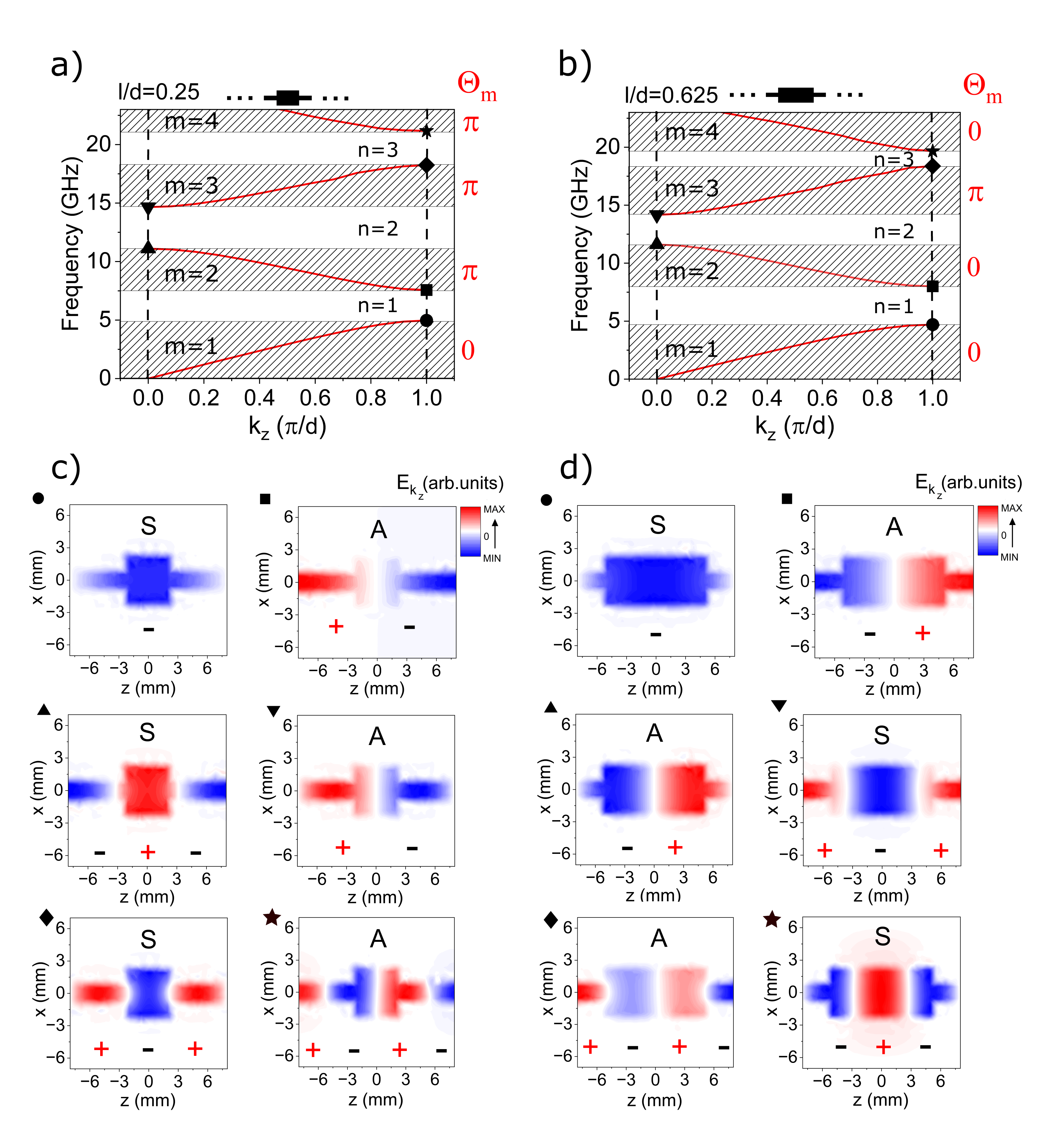}
\caption{(a,b) The numerically calculated dispersion relation for an infinite periodic microstrip, for two selected values of the bulk parameter: (a) $l/d=0.25$ and (b) $l/d=0.625$. We have kept the period $d$ constant and increased the length of the wider segment $l$ by the cost of the length of the narrower segment $d-l$ (see the black schemes in the insets above (a) and (b)). 
The values 0 and $\pi$, printed in red on the right of sub-figures (a) and (b), denote the Zak phase $\theta_m$ for successive bands $m=1,2,3,4$. (c,d) Spatial distribution of the electric field amplitude of the Bloch functions $E_{k_z}(x,z)$, for (c) $l/d=0.25$ and (d) $l/d=0.625$. The intensity of the color indicates the amplitude, while the red and blue colors indicate opposite phases. It can be seen that the symmetry of the Bloch functions at the edges of the second (up-triangle, down-triangle) and third (diamond and star) gaps are switched between symmetric (S) and anti-symmetric (A), while we change $l/d$ from (c) $0.25$ to (d) $0.625$. The systems for $d/l=0.25$ and $0.625$ are not topologically equivalent because differ in Zak phases $\theta_m$ for corresponding bands.}
\label{fig:dispersion}
\end{figure}

The exemplary dispersions $f(k_z)$ (frequency $f$ {\it versus} wave number $k_z$), computed numerically (for infinite periodic microstrip for two selected values of the relative length of the wider segment: $l/d=0.25$ and $0.625$, where the cell size is fixed at $d=16$ mm), are plotted in Fig.~\ref{fig:dispersion}(a,b). The value of $l/d=0.625$ and the values of other parameters (for both plotted dispersions) are the same as for the experimentally studied system. Although both dispersions appear similar with the frequency bands (shaded areas, which are indexed by integer number $m$) separated by band gaps (white areas, indexed with $n$), their frequency bands are not equivalent. It is worth noting that for frequencies higher than 24 GHz the dispersion relation is more complicated, as it contains branches representing, for example, the mods with nodal lines parallel to the waveguide axis.  In our work, we will limit ourselves only to mods at the lowest frequencies, where mods are quantized only along the microstrip axis (as in Fig.~\ref{fig:dispersion}).

In can be seen on the maps plotted in Fig.~\ref{fig:dispersion}(c,d), which show the profiles of the $y$-component of the electric field in a single unit cell for the frequencies from the edges of the bands/gaps (we assume the electric (magnetic) field is polarized in $y$- ($x$-) direction). The profiles are symmetric $S$ or antisymmetric $A$ while reversed in the $z$-direction. However, the symmetries of modes of some bands are different for $l/d=0.25$ and $l/d=0.625$. In particular, at the bottom (top) of the second gap -- $n=2$, the profiles of the Bloch functions are symmetric (antisymmetric) for $l/d=0.25$ and antisymmetric (symmetric) for $l/d=0.625$. This means that if we continuously change the ratio $l/d$ between the values $0.25$ and $0.625$, at some value the exchange in the mode symmetry between the bands surrounding the second gap will occur. We will see that these changes are related to the closing and reopening of the gap(s) at certain values of the ratio $l/d$ and connected with the changes of the bands Zak phases $\theta_m$ for the bands.

\begin{figure}[!ht]
\centering
\includegraphics[width=0.9\linewidth]{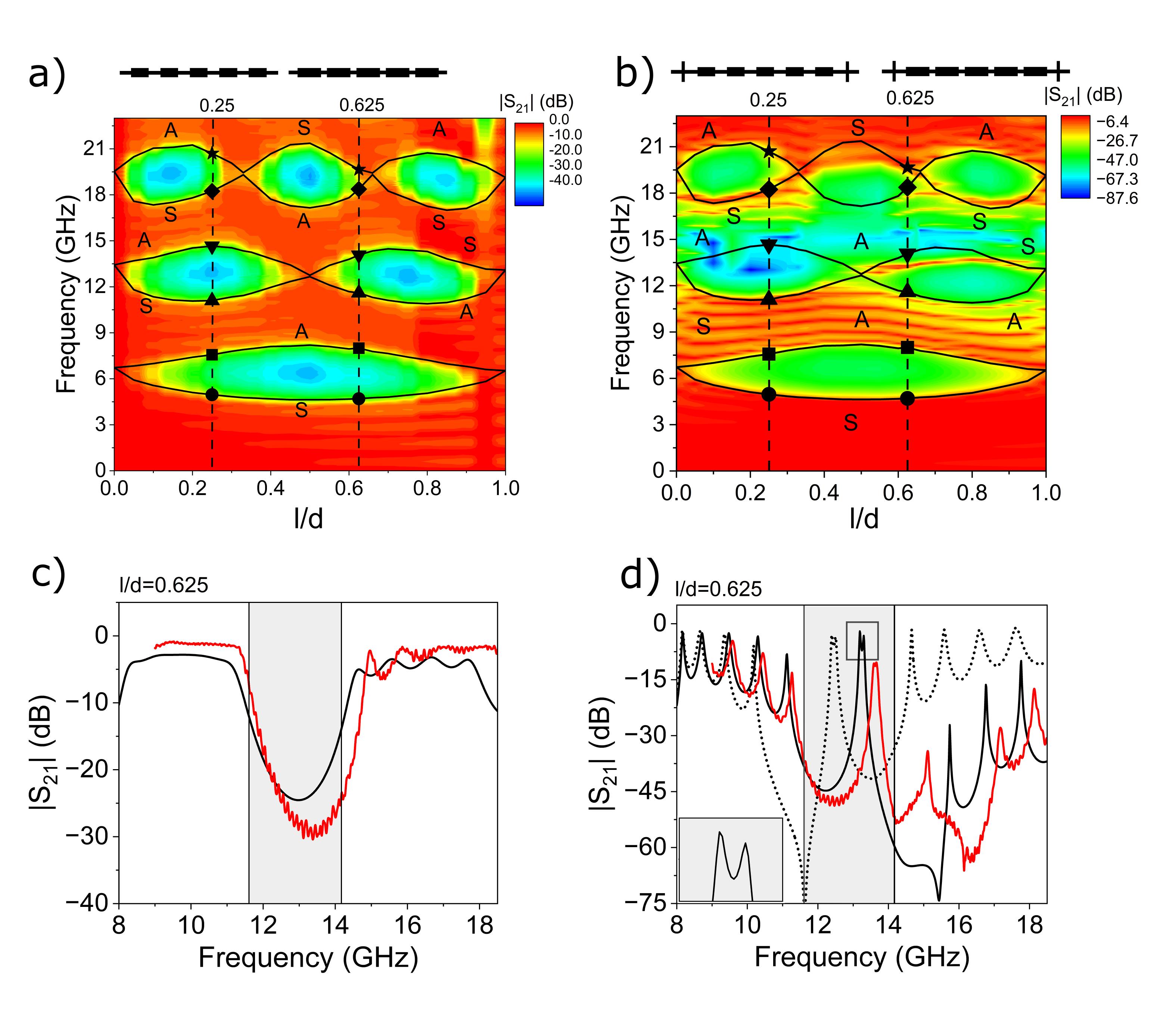}
\caption{(a,b) The 2D transmission spectra of the finite periodic microstrip numerically calculated for different values of the bulk parameter: $l/d$. The solid black lines mark the edges of the bands for the infinite microstrip, corresponding to $k_z=0$ or $k_z=\pi/d$. Two ratios $l/d=0.25$ and $0.625$, for which the exemplary dispersion relations and Bloch functions were plotted in Fig.~\ref{fig:dispersion}, are indicated by vertical dashed lines. The symmetry of the Bloch function at the edges of the band is indicated by the letters S and A. We considered the system (a) composed of five centrosymmetric cells and (b) its modification, where we added cells of modified sizes at the beginning and end of the microstrip -- see the black schemes above (a) and (b). The sizes of the edge cells and all other parameters are the same as those given in the System section. (c,d) The cross-section of the 2D spectra (a,b) at $l/d=0.625$ (solid black curves) is supplemented by the measured (red curve) transmission spectra for fabricated structures. For the microstrip with additional cells of modified sizes, we can identify the doublet transmission peaks (see enlarged view in the inset) in the second frequency gap (gray area), see their profiles in Fig.~\ref{fig:profiles}. The black dotted line in (d) shows the transmission spectra for the microstrip with modified edge cells: $d_0=12.5$ mm, $l_0=0.5$ mm, $w_0=11$ mm.
 }
\label{fig:band_structure}
\end{figure}

This effect is illustrated in Fig.~\ref{fig:band_structure}(a) showing numerically calculated transmission spectra $|S_{21}|$ in dependence on $l/d$. Here, the black lines mark the edges of the bands/gaps. We can see that in the whole $l/d$ range ($0<l/d<1$), the first gap is closed once, the second gap -- twice, the third gap -- three times, ect. It is clear that once we pass from $l/d=0.25$ (Fig.~\ref{fig:dispersion}(a)) to $l/d=0.625$ (Fig.~\ref{fig:dispersion}(b)), the order of the edges for the first (second) gap is reversed at $l/d\approx 0.50$. The topological properties of the bands associated with such a change in the spectrum of a 1D photonic crystal can be described by the Zak phase $\theta_m$ \cite{berry_quantal_1984,zak89,Wang_2019} defined for each $m^{\rm th}$ band:
\begin{equation}
    \theta_{m}=i\int_{-\pi/d}^{\pi/d}\left(\int_{\rm unit\,cell}\varepsilon(z)u_{m,k_z}^{*}(z)\partial_{k_z}u_{m,k_z}(z)dz\right)dk_z,\label{Eq:Zak}
\end{equation}
where $\varepsilon(z)$ is the effective dielectric constant taking (when $z$ is changing) two different values in wider and narrower section of microstrip (see Supplementary Information S1), and $u_{m,k_z}(z)$ is the periodic component of the Bloch function of the electric field $E_{m,k_z}(z)=u_{m,k_z}(z)e^{ik_zz}$ [$u_{m,k_z}(z)=u_{m,k_z}(z+d)$] . The Zak phase is equal to $0$ or $\pi$ for the system with centrosymmetric unit cell\cite{zak89} and can be determined by the symmetry of the Bloch function $E_{m,k_z}(z)\;$ \cite{Kohn,Zak85, Xiao14, mieszczak2022}. If the Bloch function has the same symmetry on both edges of the band (i.e. the Bloch function is symmetric (S) or antisymmetric (A) on both edges), then the Zak phase is equal to $0$, otherwise it is $\pi$. Therefore, the crossing of the edges of a given gap and the associated exchange of the gap boundaries observed during the continuous change of the bulk parameter $l/d$ leads to the exchange of the Zak phase $0\leftrightarrow\pi$.

 We observe such crossings during the transition from $d/l=0.25$ to $d/l=0.625$ in Fig.~\ref{fig:band_structure}(a). For $l/d\approx 0.5$ ( $l/d\approx 0.3$), the edges of the second (the third) gap are crossed. Therefore, the Zak phase is flipped once (at $l/d\approx 0.5$), from $\pi$ to 0, for the second band, and twice, from $\pi$ to 0 (at $l/d\approx 0.3$) and from 0 to $\pi$ again (at $l/d\approx 0.5$), for the third band. As a result the phase is the same (opposite) for the second (the third) band at $l/d=0.25$ and $0.625$ [see Fig.~\ref{fig:dispersion}(a,b) and Eq.~\ref{eq:Zak}].

\subsection*{Bulk-edge correspondence}
Moving on to the finite microstrip and determining the bandgap topology, we need to consider the boundary conditions between the periodic microstrip and the ports.
It can be analyzed with matching of the wave impedances at the ends of the periodic microstripe to the impedance of the ports. 

The spatially dependent wave impedance in an infinite periodic structure is called the Bloch impedance\cite{Collin01}:
\begin{equation}
    Z_{B}(z,f)=\frac{E_{k_z}(z,f)}{H_{k_z}(z,f)},
\end{equation}
where $E_{k_z}(z)$ and $H_{k_z}(z)$ are the Bloch waves representing the electric and magnetic fields, and $k_z(f)$ is the frequency-dependent wave number, which takes real and complex values in the frequency bands and gaps, respectively. From here on we will skip the band index $m$ in field symbols, as here the impedance can be defined for frequencies from the bands as well as the band gaps.
  The Bloch impedance can be expressed also in  terms of the reflection coefficient $r(z,f)$ and the wave impedance $Z(z,f)$ of the homogeneous sections of the structure:
\begin{equation}
    Z_{B}(z,f) \approx Z(z,f)\frac{1+r(z,f)}{1-r(z,f)}.
    \label{eq:Bloch_imped}
\end{equation}
In this case, $Z(z,f)$ corresponds to the wave impedance of the straight sections of the microstrip $Z_1$ and $Z_2$, which alternate along the structure with $z$ -- Supplementary Information S1. 
Equation (\ref{eq:Bloch_imped}) is formulated for the reflection coefficient defined as $r(z,f)=E^-_{k_z}(z,f)/E^+_{k_z}(z,f)$, where $E^-_{k_z}(z,f)$ ($E^+_{k_z}(z,f)$) are the amplitudes of the incoming (reflected) wave from the left, i.e. for the positive direction of the $z$-axis. If we consider the opposite scenario $r(z,f)=E^+_{k_z}(z,f)/E^-_{k_z}(z,f)$, which corresponds to the termination of the structure on the right side, then the sign (argument) of the Bloch impedance is reversed.  See further details in Supplementary Information S2.

For truncated systems, the Bloch impedance approaches the surface impedance of the periodic microstrip\cite{Xiao14}. If the microstrip is terminated at the edge of the centrosymmetric unit cell, then the Bloch impedance at this termination is purely imaginary for the frequencies from the band gaps, and can be written as 
\begin{equation}
Z_{B}(f)/Z_{0}=i\xi(f),
\label{Eq:gap_impedance}
\end{equation}
where $Z_0$ is the impedance of the input (output) ports, which is real and usually equal 50~$\Omega$ (see details in Supplementary Information S2). The function $\xi(f)$ changes monotonically as the frequency $f$ rises from the bottom to the top of the frequency gap, where it reaches zero or pole\cite{Zak85}. The sign of $\xi(f)$ is then constant within a given ($n^{\rm th}$) gap, and moreover it can be determined by the Zak phases $\theta_m$ of all bands below it, i.e. for $m=1,\ldots,n$:
\begin{equation}
    {\rm sign}(\xi_n)=(-1)^n\exp\left(i\sum_{m=1}^{n}\theta_m\right).\label{eq:Zak}
\end{equation}
 The formulas (\ref{eq:Zak}) and (\ref{eq:Bloch_imped}) express the bulk-edge correspondence because they relate the topological parameter of the bands (i.e., the Zak phase) to the parameter characteristic of the edge (i.e., the surface impedance). 

The impedance matching condition at the interface at a frequency $f$ is $Z_L(f)=-Z_R(f)$, where $Z_L$ and $Z_R$ are the wave impedances on the left and right sides of the interface.  In the frequency gaps, the Bloch impedance $Z_B$ taken at the edge of the centrosymmetric unit cell is imaginary (see Eq.~\ref{Eq:gap_impedance}). Therefore, $Z_B$ cannot match the real impedance $Z_0$ at the input and output ports, and thus, the edge states cannot appear in the microstrip terminated at this point.  This means that {\em the edge states in a microstrip composed of identical centrosymmetric unit cells could not exist}, independent of the topology of band structure, defined by the values of Zak phase (\ref{eq:Zak}) \cite{Zak85,Shockley39}.
{\em To induce the edge state, we need to modify the geometry of the periodic microstrip near its termination}, which will result in a change in its impedance at band gap frequencies. 
Note that the problem of the mismatch between the imaginary value of the Bloch impedance and the real value of the port impedance $Z_0$ is similar to the case of the photonic crystal\cite{Vinogradov06}, where the impedance of the vacuum (or positive epsilon medium) is also real.

The color map in Fig.~\ref{fig:band_structure}(a) shows the transmission spectrum $|S_{21}(f)|$ calculated for the periodic microstrip of finite length, depending on the bulk parameter $l/d$. The microstrip consists of five identical and symmetric cells connected to the 50~$\Omega$ ports. Therefore, the structure is terminated (i.e., the input and output ports are located) in the center of the narrower section of the unit cell. It is clear that in the frequency gaps of the infinite structure, marked by the loops surrounded by black lines, the transmission is reduced and we do not observe any peaks (lines) indicating the presence of edge modes. In the next subsection, we will study the microstrip with additional cells placed at the end and beginning of the structure, which have different sizes from the bulk cells. We will show that in such a system, we can find the edge modes. In addition, we will demonstrate the possibility of microwave tunneling that is mediated by the edge modes.

\subsection*{Edge modes}

 \begin{figure}[!ht]
\centering
\includegraphics[width=0.8\linewidth]{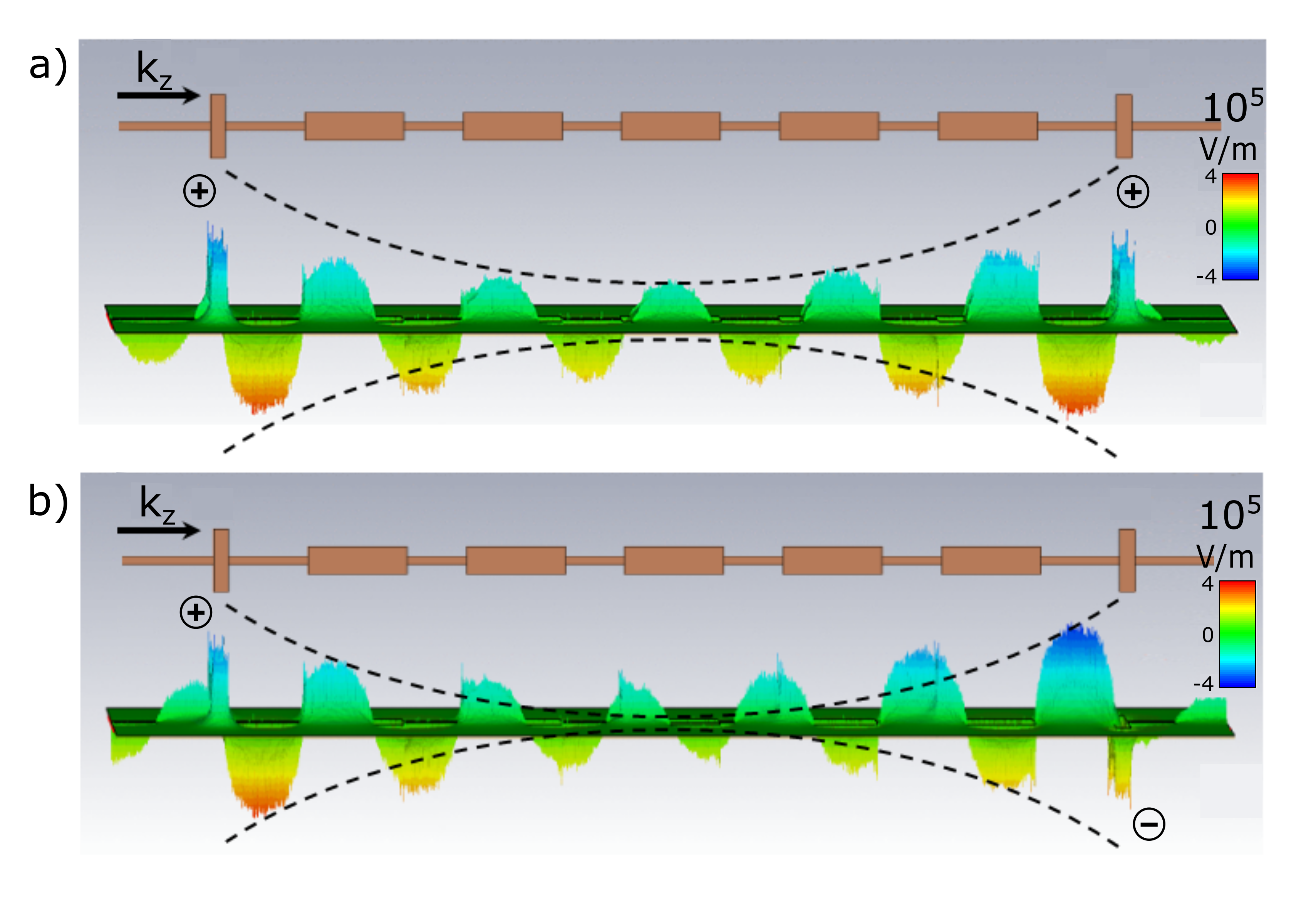}
\caption{The calculated profiles of the edge modes ($y$-component of the electric field) observed in the second gap at the frequency $~13.1$ GHz for the microstrip with additional cells of modified geometry (see Fig.~\ref{fig:band_structure}(b,d)) -- we took the parameters corresponding to the experimentally realized structure. The pair of modes of very close frequencies shows the hyperbolic decay of the amplitude, which we move away for the edges of the microstrip. The modes form a double peak in the transmission spectrum -- see Fig.~\ref{fig:band_structure}(d). The slight frequency splitting of this doublet results from the overlapping of the exponentially decaying 'tails' of the modes in the center for the structure, which is different for the mode that (a) does not flip and (b) does flip its phase (see '+' and '-' signs) as the electromagnetic wave tunnels between the edges.  
\label{fig:profiles}}
\end{figure}

 Figure~\ref{fig:band_structure}(b) shows the transmission spectrum $|S_{21}(f)|$ in dependence on $l/d$ calculated for the finite structure with additional edge cells (shown in Fig.~\ref{fig:system}(b)) located at the beginning and end of the periodic microstrip. The parameters describing the sizes of the first and last cells take the values $w_0=9$ mm, $l_0=1.5$ mm, $d_0=11.5$ mm, and $p_0=0$ (the same as for the experimentally realized system).  
 The transmission spectrum preserves the main properties of the spectrum for the unperturbed microstrip shown in Fig.~\ref{fig:system}(a). However, there are also differences: suppressing of $|S_{12}|$ in the 3$^\text{rd}$ band, and the presence of the additional line of enhanced transmission in the second gap opened for $l/d>0.5$. It enters the gap from its top edge (for $l/d$ close to 0.5) and crosses this edge again at $l/d \approx 0.9$. 
 
To gain deeper insight into obtained numerical results, we plotted (in Fig.~\ref{fig:band_structure}(c,d)) the calculated frequency-dependent transmission spectra (black solid lines) around the frequencies of the second band gap for the microstrip composed of five identical unit cells (Fig.~\ref{fig:band_structure}(c)) and this microstrip with additional edge cells (Fig.~\ref{fig:band_structure}(d)) for $d/l=0.65$ (vertical dashed line in Fig.~\ref{fig:band_structure}(a,b), i.e., the same as in experimentally realized structures. There is a visible gap in the unperturbed microstrip (shown as a gray area in Fig.~\ref{fig:band_structure}(c)), and another peak is observed within the gap at 13.1 GHz when edge cells are added to the microstrip, Fig.~\ref{fig:band_structure}(c). 
It is worth noting that the peak appears in the form of an almost degenerate doublet as is emphasized in the inset. We can identify these states as the edge modes that have a symmetric or antisymmetric profile, with respect to the center of the whole microstrip as shown in Fig.~\ref{fig:profiles}. 
  This means that  the phase of the resonantly tunneling microwaves is preserved or reversed, depending on how precisely we choose its frequency to activate one of the edge states. 
 
\begin{figure}[!ht]
\centering
\includegraphics[width=0.9\linewidth]{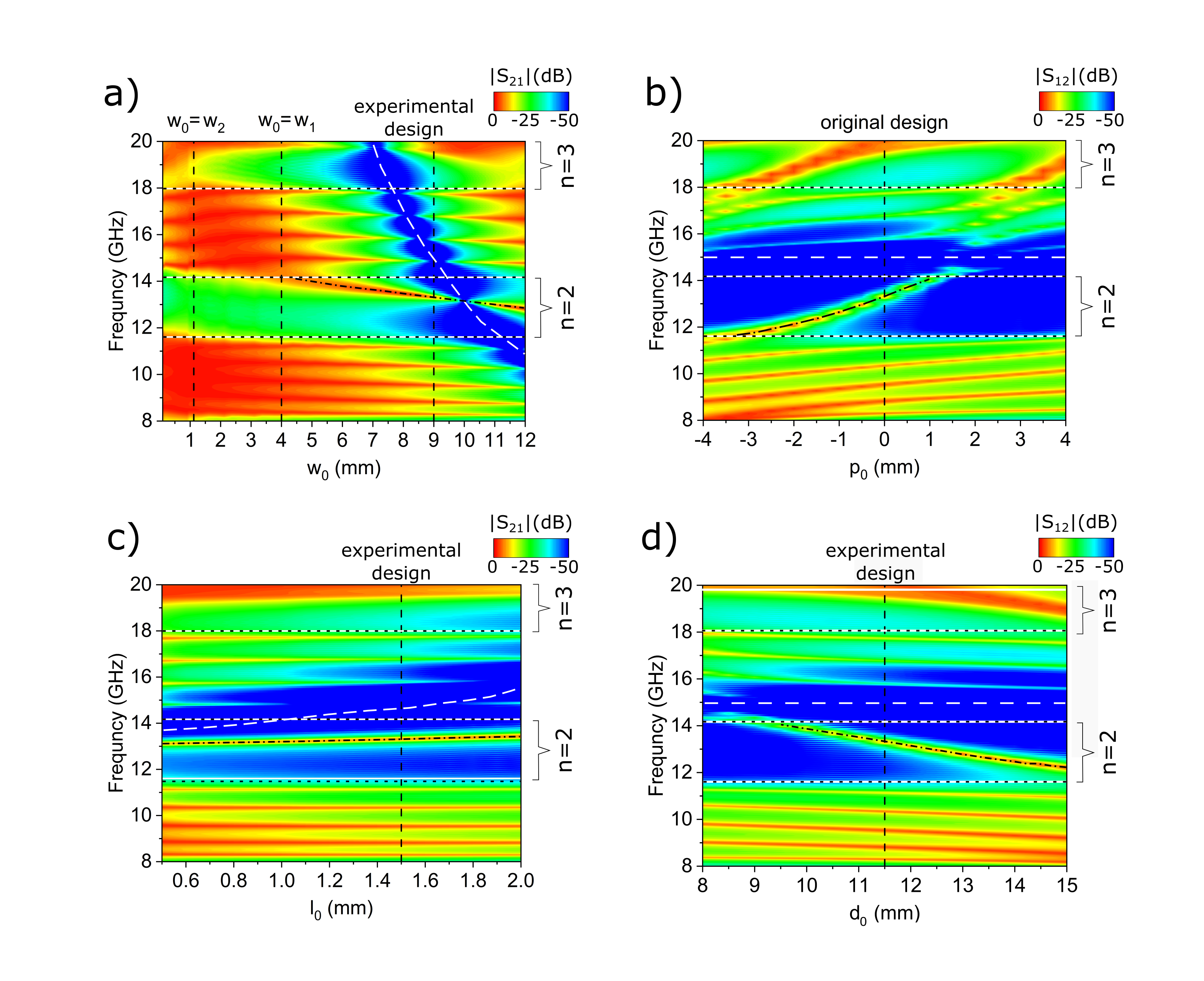}
\caption{The 2D transmission spectra of the finite periodic microstrip, calculated for different values of the parameters for modified edge cells: (a) $w_0$, (b) $p_0$, (c) $l_0$, (d) $d_0$ -- see Fig.~\ref{fig:system}(b). We have fixed the values of the bulk parameters to those corresponding to the experimental design, i.e., $l/d=0.624$; therefore, the edges of the bands (horizontal black dotted lines) are unchanged in this study. For each subplot, we change only one parameter of the edge cells. The rest corresponds to the experimental design (the vertical dashed line indicates the cases where all parameters are the same as for the fabricated structure). We can see that by adjusting the parameters: $w_0$, $p_0$ and $d_0$, one can effectively tune the frequency of the edge modes (black dashed-dotted lines). The strong attenuation in the bands is particularly high for certain values of the parameters: $w_0$ and $l_0$, when we observe the sharp minimum in the spectrum -- anti-resonance, marked by the white dashed line.}
\label{fig:edge_cell}
\end{figure}

These numerical results are superimposed with the results of measurements described in Sec. Measurements in Fig.~\ref{fig:band_structure}(c-d).  The simulations successfully replicate the majority of the features in the measured transmission spectra. However, there is a slight shift of both the frequency gap and the position of the edge states within the gap by $~0.5$ GHz. This difference can be attributed to the parameter values assumed in simulations, which may slightly differ from their nominal values in the actual sample.
In both spectra, numerical and measured, in Fig.~\ref{fig:band_structure}(d), we can see the deep anti-resonance, at 15.5 GHz (for simulations) or at 16.5 GHz (for the experiment). This anti-resonance significantly degrades the transmission through the third band. This unwanted effect is absent in the structure composed of identical cells. This means that the anti-resonance is induced by the presence of additional cells with modified sizes. However, its origin has not been elucidated yet.

As shown in the previous section, the second gap is topologically distinguishable in two ranges $l/d<0.50$ and $l/d>0.50$, separated by the point where its boundaries cross. For these two ranges, the sum of the Zak phases (for the bands below the second gap) is different (i.e. is equal to $\pi$ for $l/d<0.50$  and 0 for $l/d>0.50$) and the corresponding signs of the imaginary component of Bloch impedance are opposite -- see Eq.~\ref{eq:Bloch_imped}.  We know that no edge states are observed in any gap, regardless of their topology, in the absence of additional reshaped cells at the beginning and the end of the microstrip. However, we have seen that the addition of such cells of specific geometry has a different effect on the appearance of edge states for the gaps differing in the sign of $\xi_n$. We are dealing with such a situation for the second frequency gap, where for $l/d>0.50$ the doublet of edge states is visible in Fig.~\ref{fig:band_structure}(b) and (d).  It shows that the changes in the topology of the band structure affect the Bloch impedance and strongly modify the conditions for the impedance match in the gap, and then significantly influence the conditions for the existence of edge modes. Similarly, we can interpret the states emerging near the edges of the third gap (Fig.~\ref{fig:band_structure}(b)). One can notice lines inside the gap close to its top edge (around 20~GHz) for $l/d>0.3$ and $l/d<0.65$.

 The geometrical parameters of the edge cells also affect the frequencies of the edge modes, which can be clearly seen in Fig.~\ref{fig:band_structure}(d), where for modified values of the parameters $d_0$ (increased by 1 mm), $l_0$ (increased by 1 mm), $w_0$ (decreased by 2 mm) the frequencies of the edge modes were shifted by 1~GHz (black dotted line). In order to study this problem systematically, we examined the effect of the parameters: $w_0$, $l_0$, $d_0$ on the transmission spectrum. In addition, we studied the role of the shift of the wider segment of the edge cell with respect to its center, i.e. we changed the parameter $p_0$.  Fig.~\ref{fig:edge_cell} shows the calculated transmission spectrum for the finite microstrip where the chain of five identical bulk cells is supplemented by edge cells of modified sizes. We kept the sizes of the bulk cells the same as in the experiment ($d=16$ mm, $l= 10$ mm, $w_1= 4$ mm, $w_2=1.12$ mm) and changed only one of the geometrical parameters of the edge cell (all other parameters of the edge cell were the same as in the experimentally realized system: $l_0=1.5$~mm, $d_0=11.5$~mm, $w_0=9$~mm, $p_0=0$). Similar to Fig.~\ref{fig:band_structure}, we plot also the boundaries of the gaps of the infinite microstrip. These boundaries are indicated by horizontal black dotted lines and allow us to identify edge modes, whenever a transmission peak enters the gaps. It is worth noting that if we change only the geometry of the edge cell and do not modify any of the bulk parameters, the edges of the gaps remain unaffected. 

Fig.~\ref{fig:edge_cell}(a) and (c) show the transmission spectrum as a function of the width and length of the central part of the edge cell: $w_0$ and $l_0$, respectively. By increasing the width $w_0$, we can see entering the edge modes from the third band into the second gap ($n=2$) below at $w_0=4$ mm, which reaches the middle of the gap at $w_0=12$ mm. The transmission is strongly suppressed by the anti-resonance (marked with the white dashed line), which enters the third gap at $w_0=7.2$ mm. The frequency of the anti-resonance decreases rapidly with increasing $w_0$. It deteriorates the transmission in third band  to very low values (below -50~dB), and also  the transmission by edge modes in second gap ($n=2$), for $w_0=10$ mm.  When we set the width to $w_0=9$~nm and start detuning $l_0$ from the experimentally chosen value of 1.5 mm (Fig.~\ref{fig:edge_cell}(c)). Surprisingly, we do not see significant changes in the frequency of the edge mode, only its slight linear increase from 13 to 23.5~GHz with the change of $l_0$ from 0.5 to 2.0 mm. However, with the changes of $l_0$ we observe a  frequency increase of the anti-resonance. 

Figure ~\ref{fig:edge_cell}(b,d) shows a scenario where we do not change the dimensions of the central part of the edge cell (width $w_0$ and length $l_0$), but move it within the edge cells by adjusting $p_0$ (b), and change the size of the edge cell itself by changing $d_0$ (d). By adjusting these two parameters we can pull the edge modes across the gap. In particular, the edge modes enter the second gap ($n=2$) from the bottom at $p_0=-4$~mm and exit at $p_0=1$~mm. In the case of $d_0$ changes, the frequency of the edge modes enters the gap $d_0=9.5$ nm decrease monotonically frequency with increasing $d_0$. Interestingly, the frequency of the anti-resonance is almost not affected by either $p_0$ or $d_0$. 

Resuming, the dependencies illustrated in Figs.~\ref{fig:edge_cell}(a)-(d) demonstrate that the frequency of the edge modes in the second gap can be modified by adjusting the shape of edge cells: $w_0$, $p_0$, and $d_0$, whereas the frequency of the anti-resonance can be controlled by $w_0$ and $l_0$. This suggests that the modified edge cell of the microstrip is responsible for both effects, although the underlying physics differs. The independent design of both can be advantageous for various applications since significant transmission suppression is observed around the anti-resonance frequency, exceeding the $\pm1$ GHz range.

Figure~\ref{fig:band_structure}(d) illustrates a doublet peak in transmission through the edge modes, consisting of symmetric and antisymmetric modes (Fig.~\ref{fig:profiles}). While the frequency splitting between these modes is minute and barely discernible in the numerical simulations presented in Fig.~\ref{fig:band_structure}, we utilise the lumped-element method to obtain further insight into the splitting of the edge modes. This model provide the relation between the  real system (microstrip) and  it simplified but useful description in the form electrical network.

\subsubsection*{Edge modes in lumped-element model}

\begin{figure}[!ht]
\centering
\includegraphics[width=0.75\linewidth]{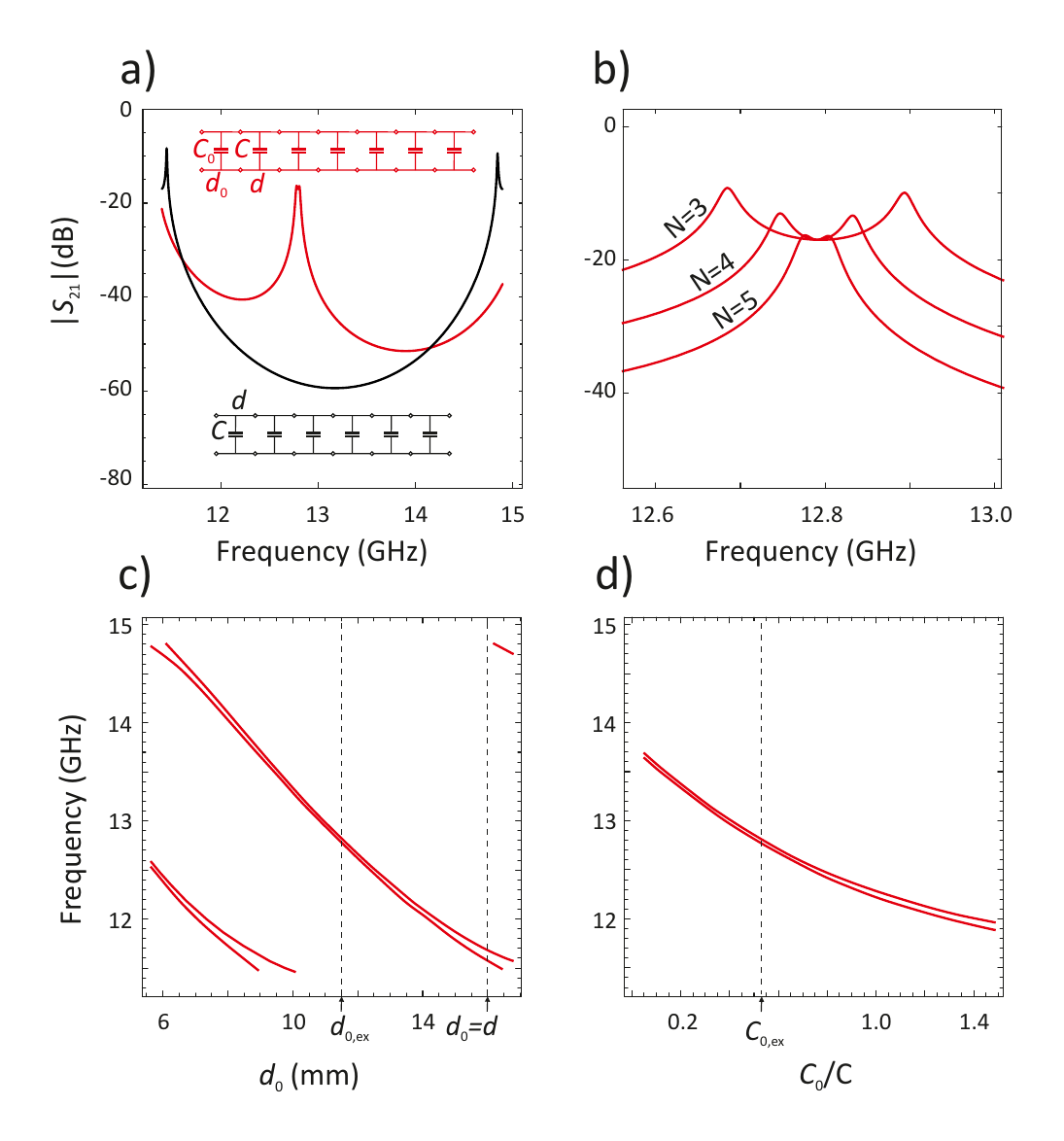}
\caption{Results for the lumped-element model. (a) The transmission (b,c) spectrum within the second frequency gap for a ladder network composed of five identical cells (black line) and extended by two cells with different parameters: $d_0\ne d$ and $C_0\ne C$ (red line). (b) The magnified peak for the doublet of edge modes become split when the number of bulk cells $N$ is reduced.  (c,d) Tuning the edge mode frequency.  We studied the effect of the length of the edge cell $d_0$ (c) and the shunt capacitance $c_0$ (d). In (b) we fixed the shunt capacitance $C_0 = 0.53 C$, and in (d) -- the length $d_0= 0.73 d$ for the edge cells. 
\label{Fig:lump_element}}
\end{figure}

The periodic microstrip is a transmission line that can be approximated by a lumped-element model as a ladder network (see Fig.~\ref{fig:system}(c)). Each unit cell of the structures can be described as a phase shifter with shunt capacitance. The phase shift depends on the length of the unit cell $d$ and the wave vector $k_0 = 2\pi f/ (c/\sqrt{\varepsilon_{\rm eff}})$ in the medium characterized by the dielectric constant $\varepsilon_{\rm eff}$, where the symbol $c$ denotes the speed of light. The shunt capacitance $C$ is related to the capacitance of a single section of microstrip deposited on a dielectric substrate with a ground plane at the bottom. The formal relationship between output and input voltages ($V_j$) and currents ($I_j$) for each segment ($j^{\rm th}$) of the network is described by the so-called ABCD-matrix\cite{Pozar11}, with frequency-dependent elements (see Supplementary Information S3). In the ladder network, the voltages and currents in successive cells are related by the Bloch wave number $k_z$: $V_j=e^{ik_zd}V_{j+1}$, $I_j=e^{ik_zd}I_{j+1}$. This allowed us to find the dispersion relation $f(k_z)$ (for infinite network) and the transfer coefficients $S_{21}$ (for finite network). To relate the lumped-element model to the full electromagnetic model used in numerical calculations, we adjusted the values of two parameters $C$ and $\varepsilon_{\rm eff}$. The accuracy of the lumped-element model is limited, and we could not properly reproduce the spectrum in a wide frequency range, covering many bands and gaps.  In this study, we matched the position and width of the second frequency gaps of infinite ladder network and periodic microstrip for the values $C=0.48$~pF and $\varepsilon_{\rm eff}=1.6$. 

The geometric changes of the edge cells in finite systems are introduced by varying the parameters: $C_0$ and $d_0$. The length of the edge cells $d_0$ is clearly defined ($d_0=11.5$ mm for the experimental realization).  We have estimated the value of $C_0$ assuming that the ratio of the area of the edge cell to the area of the bulk cell: $S_0/S$ can be related to the corresponding ratio of capacitances: $S_0/S \approx C_0/C$.  By calculating the transmission for the ladder network, we found that the transmission peak is located inside the gap, close to the experimentally determined frequency -- see Fig.~\ref{Fig:lump_element}(a). Careful inspection allows us to note that the peak is doubled. 

In Fig.~\ref{Fig:lump_element}(b,c) we show the dependence of the edge mode frequencies on $d_0$ and $C_0$ in the second frequency gap for the lumped-element model. By changing the length of the edge cell $d_0$ from 6 to 19 mm, we shift the edge mode doublet from the top to the bottom of the frequency gap -- Fig.~\ref{Fig:lump_element}(b). The doublet increases the splitting near the edges of the frequency gap. This is due to the imaginary part of the wavenumber, which has a maximum value at the center of the gap and decreases as one moves toward the edges. Close to the frequency band, the surface modes are less localized and the excitations at both edges of the microstip are strongly coupled. The lumped-element model predicts that for small values of $d_0$ (less than 10 mm) an additional doublet of surface modes appears at the bottom of the frequency gap. However, such states were not found using numerical simulations (Fig.~\ref{fig:edge_cell}(c)) in the investigated range of $d_0$. We assume that this is an artifact introduced by the lumped-element model.
 
 The frequency of edge modes can also be tuned by the shunt capacitance $C_0$ in the lumped element model (see Fig.~\ref{Fig:lump_element}(d)). The increase of the capacitance for the edge cells of the network can be related to the widening of the central segment in the edge cells of the microstrip. In both models, the frequency of the edge modes is reduced with the increase of $C_0$ ($w_0$).

 The frequency difference between the two edge modes decreases as the number of cells increases -- see Fig.~\ref{Fig:lump_element}(b). This can be understood by noting that for a larger number of cells $N>5$, the amplitudes of both modes practically decay to zero inside the periodic microstrip. As a result, the microwave fields at both edges do not affect each other. For such systems, it is very difficult to choose the frequency that allows microwave-field tunneling with 0 or $\pi$ phase shift between input and output port (see Fig.~\ref{fig:profiles})

 \section*{Discussion}

We have studied experimentally and numerically the transmission of microwaves through a microstrip of periodically modulated width. We have shown that 5 repetitions of the unit cell are sufficient to observe the band gap in the transmission spectrum. We show that this gap has topological property allowing for the existence of the edge modes. 

However, in the undefected microstrip (composed of centrosymmetric unit cells), the edge states cannot exist and the modification of the first and last cell is required to induce them, which was demonstrated experimentally. This effect was explained by relating the topology of the 1D band structure, i.e., the Zak phases of the bands, to the wave impedance of the microstrip (Bloch impedance).  The amplitudes of these modes are located at both ends of the microstrip, enabling the passage of microwaves throughout the entire structure. We can manipulate the edge modes within the desired frequency gap by modifying the geometric parameters of the first and last cell. Nevertheless, the anti-resonance originating in the edge unit cells also impacts the transmission of the microstrip. Thus, using the microstrip for applications such as filtering necessitates careful design of both edge modes and anti-resonance. Fortunately, the geometric parameters of the microstrip's edge cells have different dependencies on these components.

The edge states appeared in the form of a narrow doublet, where the first (second) mode does (does not) flip its phase while being transmitted through the structure. Therefore, the investigated system can be considered as a narrow passband filter of high selectivity. The phase delay in the passband of such a filter is significant because the phase shift between output and input changes from 0 to $\pi$ when the frequency is tuned between the peaks in the doublet.  


\section*{Methods}

\subsection*{Numerical studies}
For the considered periodic microstrips, the transmission spectra $S_{21}$ as well as dispersion diagrams and phase distribution maps were calculated using {\em CST Microwave Studio}. For the simulation of the dispersion diagrams, the eigenmode calculation with periodic boundary conditions (PBC) in the $z-$ direction was used. The PBC allows modeling an infinite extension of the proposed bulk cell. The "Phase" parameter in PCB is chosen to sweep from 0 to 180 degrees in equal steps, which allows to change the Bloch phase and the associated wave vector from the center to the edge of the $1^{\rm st}$ Brillouin zone. The boundary conditions along the $x-$ and $y-$ axes are equivalent to an introduction to the perfect electrically conducting (PEC) plane. The frequency range starts at 0 Hz and ends at 27 GHz. The Eigenmode Solver has been set to calculate a certain number of the lowest resonant frequencies of the structure, since only the fundamental modes are of interest.  The Frequency Domain Solver with tetrahedral mesh is used to calculate the S-parameters and the transmission spectra. The following calculation requires the definition of waveguide ports in $z-$ direction through which energy enters and leaves the structure. The open boundary implementation of the solver is the recommended setting. Open boundaries for this type of simulation are implemented as Floquet (Bloch) mode ports and realized in $x-$ and $y-$ planes. 

For the lumped element model, the calculation of the transmission spectra $S_{21}$ and the frequencies of the edge modes was done in {\em Mathematica 12}. For each cell we defined the ABDC matrix\cite{Pozar11}. The product of these matrices for the sequence of cells, in finite microstrip, was used to find the S-matrix and the transmission spectrum. The details of these calculations are presented in Supplementary Information S3.

\subsection*{Experimental studies}
We used the vector network analyzer (VNA), Agilent N5230A PNA-L, to measure the transmission spectra of the microwaves in the frequency range of 8 - 20 GHz (see the photo of the experimental setup). The input and output of the microstrip are connected to a vector network analyzer with standard 50 $\Omega$ ports to measure the $S_{21}$ spectrum. The unit cell consists of sequentially connected narrow and wide segments. The periodic stripes are etched by photolithography on one side of the Neltec substrate, which is metallized underneath, so that the line consists of five cells and two edge cells.  The geometric parameters at the ends of the structure correspond to an input impedance of $Z_0 = 50$ $\Omega$. The microwave signal is emitted from one port, transmitted through the structure, and received at another port.

\section*{Data availability}
Data supporting this study are openly available from the repository at https://zenodo.org/records/10203785.



\section*{Acknowledgements }
This work was supported by NCN Poland project no. UMO-2020/39/I/ST3/02413 and the European Union’s Horizon 2020 research and innovation programme under the Marie Skłodowska-Curie grant agreement No 644348 (MagIC), and the Polish agency NAWA under the grant  No PPN/BUA/2019/1/00114. J.W.K. would like to acknowledge the support of the Foundation of Alfried Krupp Kolleg Greifswald. The authors would like to thank Andriy E. Serebryannikov for fruitful discussion,  are grateful to M. Baranowski and S. Mieszczak for their contributions in the early stages of this work.

\section*{Author contributions statement}
S.T. designed the experiments, A.G. and G.K. performed measurements of microwave frequency spectra, A.G. fabricated the microstrip under study. L.I. and S.P. performed the numerical simulations. M.K. and J.W.K. conducted the theoretical analysis. L.I. provided the description of the methods section. J.W.K. provided the description of the lumped element model and performed the lumped element calculations. All authors contributed to the manuscript. 

\section*{Ethics declarations}
\subsection*{Competing interests} 
The authors declare no competing interests.

\clearpage
\includepdf[pages=1]{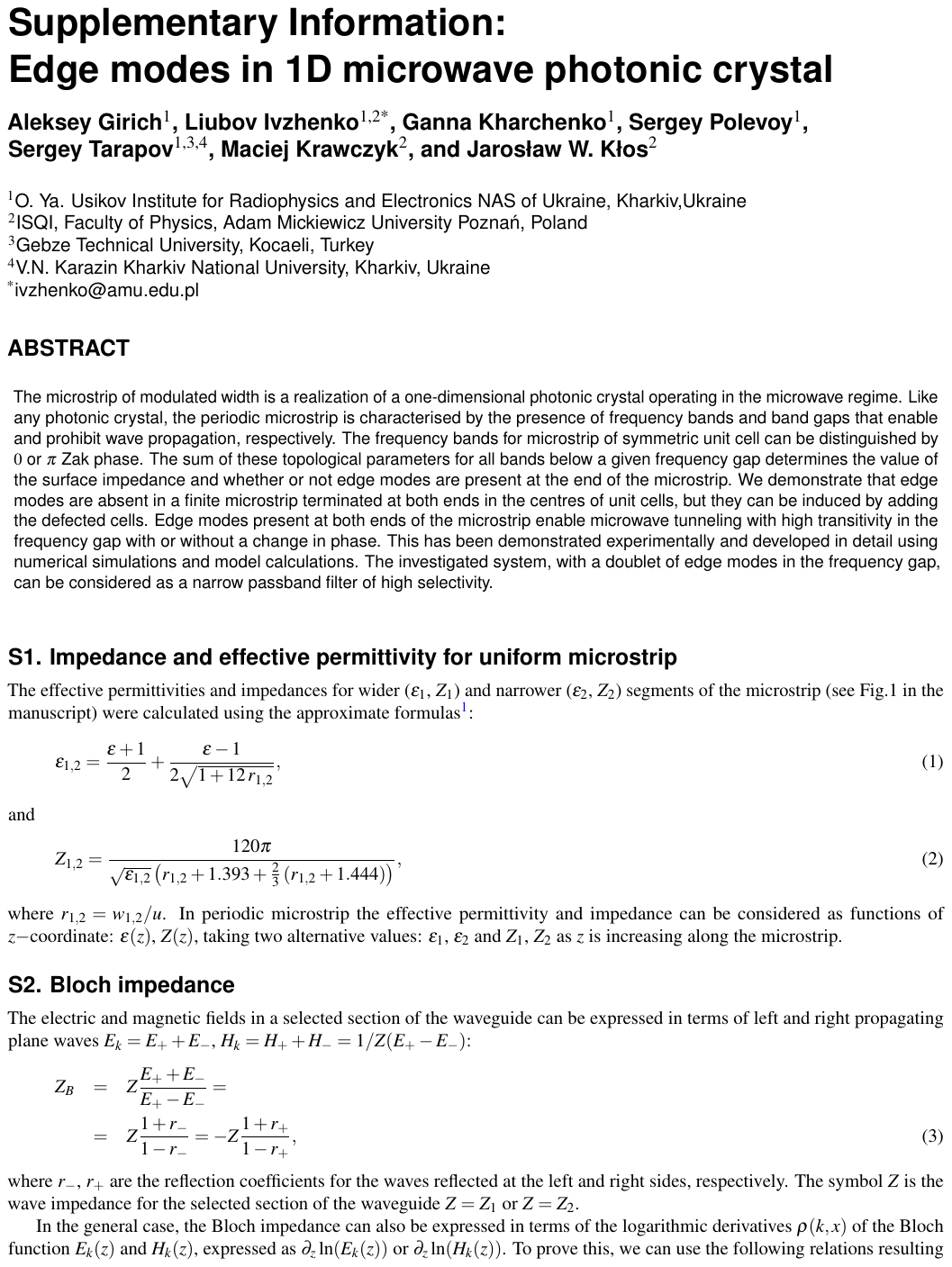}
\clearpage
\includepdf[pages=2]{SI_final.pdf}


\begin{thebibliography}{10}
\urlstyle{rm}
\expandafter\ifx\csname url\endcsname\relax
  \def\url#1{\texttt{#1}}\fi
\expandafter\ifx\csname urlprefix\endcsname\relax\def\urlprefix{URL }\fi
\expandafter\ifx\csname doiprefix\endcsname\relax\def\doiprefix{DOI: }\fi
\providecommand{\bibinfo}[2]{#2}
\providecommand{\eprint}[2][]{\url{#2}}

\bibitem{Zak82}
\bibinfo{author}{Zak, J.}
\newblock \bibinfo{journal}{\bibinfo{title}{Band center---a conserved quantity
  in solids}}.
\newblock {\emph{\JournalTitle{Phys. Rev. Lett.}}}
  \textbf{\bibinfo{volume}{48}}, \bibinfo{pages}{359--362},
  \doiprefix\url{10.1103/PhysRevLett.48.359} (\bibinfo{year}{1982}).

\bibitem{zak89}
\bibinfo{author}{Zak, J.}
\newblock \bibinfo{journal}{\bibinfo{title}{Berry's phase for energy bands in
  solids}}.
\newblock {\emph{\JournalTitle{Phys. Rev. Lett.}}}
  \textbf{\bibinfo{volume}{62}}, \bibinfo{pages}{2747--2750},
  \doiprefix\url{10.1103/PhysRevLett.62.2747} (\bibinfo{year}{1989}).

\bibitem{atala_direct_2013}
\bibinfo{author}{Atala, M.} \emph{et~al.}
\newblock \bibinfo{journal}{\bibinfo{title}{Direct measurement of the {Zak}
  phase in topological {Bloch} bands}}.
\newblock {\emph{\JournalTitle{Nature Physics}}} \textbf{\bibinfo{volume}{9}},
  \bibinfo{pages}{795--800}, \doiprefix\url{10.1038/nphys2790}
  (\bibinfo{year}{2013}).

\bibitem{Xiao14}
\bibinfo{author}{Xiao, M.}, \bibinfo{author}{Zhang, Z.~Q.} \&
  \bibinfo{author}{Chan, C.~T.}
\newblock \bibinfo{journal}{\bibinfo{title}{Surface impedance and bulk band
  geometric phases in one-dimensional systems}}.
\newblock {\emph{\JournalTitle{Phys. Rev. X}}} \textbf{\bibinfo{volume}{4}},
  \bibinfo{pages}{021017}, \doiprefix\url{10.1103/PhysRevLett.48.359}
  (\bibinfo{year}{2014}).

\bibitem{Yeh_77}
\bibinfo{author}{Yeh, P.}, \bibinfo{author}{Yariv, A.} \&
  \bibinfo{author}{Hong, C.-S.}
\newblock \bibinfo{journal}{\bibinfo{title}{Electromagnetic propagation in
  periodic stratified media. {I.} {G}eneral theory}}.
\newblock {\emph{\JournalTitle{J. Opt. Soc. Am.}}}
  \textbf{\bibinfo{volume}{67}}, \bibinfo{pages}{423--438},
  \doiprefix\url{10.1364/JOSA.67.000423} (\bibinfo{year}{1977}).

\bibitem{Vinogradov06}
\bibinfo{author}{Vinogradov, A.~P.} \emph{et~al.}
\newblock \bibinfo{journal}{\bibinfo{title}{Surface state peculiarities in
  one-dimensional photonic crystal interfaces}}.
\newblock {\emph{\JournalTitle{Phys. Rev. B}}} \textbf{\bibinfo{volume}{74}},
  \bibinfo{pages}{045128}, \doiprefix\url{10.1103/PhysRevB.74.045128}
  (\bibinfo{year}{2006}).

\bibitem{Klos07}
\bibinfo{author}{K{\l}os, J.}
\newblock \bibinfo{journal}{\bibinfo{title}{Conditions of {T}amm and {S}hockley
  state existence in chains of resonant cavities in a photonic crystal}}.
\newblock {\emph{\JournalTitle{Phys. Rev. B}}} \textbf{\bibinfo{volume}{76}},
  \bibinfo{pages}{165125}, \doiprefix\url{10.1103/PhysRevB.76.165125}
  (\bibinfo{year}{2007}).

\bibitem{Malkova09}
\bibinfo{author}{Malkova, N.}, \bibinfo{author}{Hromada, I.},
  \bibinfo{author}{Wang, X.}, \bibinfo{author}{Bryant, G.} \&
  \bibinfo{author}{Chen, Z.}
\newblock \bibinfo{journal}{\bibinfo{title}{Transition between tamm-like and
  shockley-like surface states in optically induced photonic superlattices}}.
\newblock {\emph{\JournalTitle{Phys. Rev. A}}} \textbf{\bibinfo{volume}{80}},
  \bibinfo{pages}{043806}, \doiprefix\url{10.1103/PhysRevA.80.043806}
  (\bibinfo{year}{2009}).

\bibitem{Wang16}
\bibinfo{author}{Wang, Q.}, \bibinfo{author}{Xiao, M.}, \bibinfo{author}{Liu,
  H.}, \bibinfo{author}{Zhu, S.} \& \bibinfo{author}{Chan, C.~T.}
\newblock \bibinfo{journal}{\bibinfo{title}{Measurement of the {Z}ak phase of
  photonic bands through the interface states of a metasurface/photonic
  crystal}}.
\newblock {\emph{\JournalTitle{Phys. Rev. B}}} \textbf{\bibinfo{volume}{93}},
  \bibinfo{pages}{041415}, \doiprefix\url{10.1103/PhysRevB.93.041415}
  (\bibinfo{year}{2016}).

\bibitem{Yilmaz18}
\bibinfo{author}{Yilmaz, D.}, \bibinfo{author}{Yeltik, A.} \&
  \bibinfo{author}{Kurt, H.}
\newblock \bibinfo{journal}{\bibinfo{title}{Highly controlled {B}loch wave
  propagation in surfaces with broken symmetry}}.
\newblock {\emph{\JournalTitle{Opt. Lett.}}} \textbf{\bibinfo{volume}{43}},
  \bibinfo{pages}{2660}, \doiprefix\url{10.1364/OL.43.002660}
  (\bibinfo{year}{2018}).

\bibitem{Wang_2019}
\bibinfo{author}{Wang, H.-X.}, \bibinfo{author}{Guo, G.-Y.} \&
  \bibinfo{author}{Jiang, J.-H.}
\newblock \bibinfo{journal}{\bibinfo{title}{Band topology in classical waves:
  Wilson-loop approach to topological numbers and fragile topology}}.
\newblock {\emph{\JournalTitle{New J. Phys.}}} \textbf{\bibinfo{volume}{21}},
  \bibinfo{pages}{093029}, \doiprefix\url{10.1088/1367-2630/ab3f71}
  (\bibinfo{year}{2019}).

\bibitem{Henriques}
\bibinfo{author}{Henriques, J. C.~G.}, \bibinfo{author}{Rappoport, T.~G.},
  \bibinfo{author}{Bludov, Y.~V.}, \bibinfo{author}{Vasilevskiy Y.~B, M.~I.} \&
  \bibinfo{author}{Peres, N. M.~R.}
\newblock \bibinfo{journal}{\bibinfo{title}{Topological photonic {T}amm states
  and the {S}u-{S}chrieffer-{H}eeger model}}.
\newblock {\emph{\JournalTitle{Phys. Rev. A}}} \textbf{\bibinfo{volume}{101}},
  \bibinfo{pages}{043811}, \doiprefix\url{10.1103/PhysRevA.101.043811}
  (\bibinfo{year}{2020}).

\bibitem{xiao_geometric_2015}
\bibinfo{author}{Xiao, M.} \emph{et~al.}
\newblock \bibinfo{journal}{\bibinfo{title}{Geometric phase and band inversion
  in periodic acoustic systems}}.
\newblock {\emph{\JournalTitle{Nature Physics}}} \textbf{\bibinfo{volume}{11}},
  \bibinfo{pages}{240--244}, \doiprefix\url{10.1038/nphys3228}
  (\bibinfo{year}{2015}).

\bibitem{yin_band_2018}
\bibinfo{author}{Yin, J.} \emph{et~al.}
\newblock \bibinfo{journal}{\bibinfo{title}{Band transition and topological
  interface modes in {1D} elastic phononic crystals}}.
\newblock {\emph{\JournalTitle{Sci. Rep.}}} \textbf{\bibinfo{volume}{8}},
  \bibinfo{pages}{6806}, \doiprefix\url{10.1038/s41598-018-24952-5}
  (\bibinfo{year}{2018}).

\bibitem{Li}
\bibinfo{author}{Li, Z.-w.}, \bibinfo{author}{Fang, X.-s.},
  \bibinfo{author}{Liang, B.}, \bibinfo{author}{Li, Y.} \&
  \bibinfo{author}{Cheng, J.-c.}
\newblock \bibinfo{journal}{\bibinfo{title}{Topological interface states in the
  low-frequency band gap of one-dimensional phononic crystals}}.
\newblock {\emph{\JournalTitle{Phys. Rev. Applied}}}
  \textbf{\bibinfo{volume}{14}}, \bibinfo{pages}{054028},
  \doiprefix\url{10.1103/PhysRevApplied.14.054028} (\bibinfo{year}{2020}).

\bibitem{wang_zak_2018}
\bibinfo{author}{Wang, L.}, \bibinfo{author}{Cai, W.}, \bibinfo{author}{Bie,
  M.}, \bibinfo{author}{Zhang, X.} \& \bibinfo{author}{Xu, J.}
\newblock \bibinfo{journal}{\bibinfo{title}{{Z}ak phase and topological
  plasmonic {Tamm} states in one-dimensional plasmonic crystals}}.
\newblock {\emph{\JournalTitle{Opt. Express}}} \textbf{\bibinfo{volume}{26}},
  \bibinfo{pages}{28963--28975}, \doiprefix\url{10.1364/OE.26.028963}
  (\bibinfo{year}{2018}).

\bibitem{poli_selective_2015}
\bibinfo{author}{Poli, C.}, \bibinfo{author}{Bellec, M.},
  \bibinfo{author}{Kuhl, U.}, \bibinfo{author}{Mortessagne, F.} \&
  \bibinfo{author}{Schomerus, H.}
\newblock \bibinfo{journal}{\bibinfo{title}{Selective enhancement of
  topologically induced interface states in a dielectric resonator chain}}.
\newblock {\emph{\JournalTitle{Nature Communications}}}
  \textbf{\bibinfo{volume}{6}}, \bibinfo{pages}{6710},
  \doiprefix\url{10.1038/ncomms7710} (\bibinfo{year}{2015}).

\bibitem{Rychly17}
\bibinfo{author}{Rychły, J.} \& \bibinfo{author}{K{\l}os, J.~W.}
\newblock \bibinfo{journal}{\bibinfo{title}{Spin wave surface states in {1D}
  planar magnonic crystals}}.
\newblock {\emph{\JournalTitle{J. Phys. D: Appl. Phys.}}}
  \textbf{\bibinfo{volume}{50}}, \bibinfo{pages}{164004},
  \doiprefix\url{10.1088/1361-6463/aa5ae1} (\bibinfo{year}{2017}).

\bibitem{Kucharczyk}
\bibinfo{author}{Ste\ifmmode \mbox{\c{}}\else \c{}\fi{}\ifmmode~\acute{s}\else
  \'{s}\fi{}licka, M.}, \bibinfo{author}{Kucharczyk, R.} \&
  \bibinfo{author}{Glasser, M.~L.}
\newblock \bibinfo{journal}{\bibinfo{title}{Surface states in superlattices}}.
\newblock {\emph{\JournalTitle{Phys. Rev. B}}} \textbf{\bibinfo{volume}{42}},
  \bibinfo{pages}{1458--1461}, \doiprefix\url{10.1103/PhysRevB.42.1458}
  (\bibinfo{year}{1990}).

\bibitem{Klos03}
\bibinfo{author}{K{\l}os, J.} \& \bibinfo{author}{Puszkarski, H.}
\newblock \bibinfo{journal}{\bibinfo{title}{Conditions of coexistence of {T}amm
  and {S}hockley states in a superlattice with a perturbed surface}}.
\newblock {\emph{\JournalTitle{Phys. Rev. B}}} \textbf{\bibinfo{volume}{68}},
  \bibinfo{pages}{045316}, \doiprefix\url{10.1103/PhysRevB.68.045316}
  (\bibinfo{year}{2003}).

\bibitem{marpaung_integrated_2019}
\bibinfo{author}{Marpaung, D.}, \bibinfo{author}{Yao, J.} \&
  \bibinfo{author}{Capmany, J.}
\newblock \bibinfo{journal}{\bibinfo{title}{Integrated microwave photonics}}.
\newblock {\emph{\JournalTitle{Nature Photonics}}}
  \textbf{\bibinfo{volume}{13}}, \bibinfo{pages}{80--90},
  \doiprefix\url{10.1038/s41566-018-0310-5} (\bibinfo{year}{2019}).

\bibitem{Kee99}
\bibinfo{author}{Kee, C.-S.} \emph{et~al.}
\newblock \bibinfo{journal}{\bibinfo{title}{Essential parameter in the
  formation of photonic band gaps}}.
\newblock {\emph{\JournalTitle{Phys. Rev. E}}} \textbf{\bibinfo{volume}{59}},
  \bibinfo{pages}{4695--4698}, \doiprefix\url{10.1103/PhysRevE.59.4695}
  (\bibinfo{year}{1999}).

\bibitem{Tarapov12}
\bibinfo{author}{Tarapov, S.~I.}
\newblock \bibinfo{journal}{\bibinfo{title}{Microwaves in dispersive magnetic
  composite media}}.
\newblock {\emph{\JournalTitle{Low Temp. Phys.}}}
  \textbf{\bibinfo{volume}{38}}, \bibinfo{pages}{603},
  \doiprefix\url{10.1063/1.4733684} (\bibinfo{year}{2012}).

\bibitem{Chernovtsev_2007}
\bibinfo{author}{Chernovtsev, S.~V.}, \bibinfo{author}{Belozorov, D.~P.} \&
  \bibinfo{author}{Tarapov, S.~I.}
\newblock \bibinfo{journal}{\bibinfo{title}{Magnetically controllable {1D}
  magnetophotonic crystal in millimetre wavelength band}}.
\newblock {\emph{\JournalTitle{J. Phys. D: Appl. Phys.}}}
  \textbf{\bibinfo{volume}{40}}, \bibinfo{pages}{295--299},
  \doiprefix\url{10.1088/0022-3727/40/2/001} (\bibinfo{year}{2007}).

\bibitem{Zhu}
\bibinfo{author}{Zhu, W.} \emph{et~al.}
\newblock \bibinfo{journal}{\bibinfo{title}{{Z}ak phase and band inversion in
  dimerized one-dimensional locally resonant metamaterials}}.
\newblock {\emph{\JournalTitle{Phys. Rev. B}}} \textbf{\bibinfo{volume}{97}},
  \bibinfo{pages}{195307}, \doiprefix\url{10.1103/PhysRevB.97.195307}
  (\bibinfo{year}{2018}).

\bibitem{wu_kind_2023}
\bibinfo{author}{Wu, C.~H.} \emph{et~al.}
\newblock \bibinfo{journal}{\bibinfo{title}{A kind of planar waveguides for
  cheating the high-speed digital signals into misidentifying the
  characteristic impedance}}.
\newblock {\emph{\JournalTitle{Scientific Reports}}}
  \textbf{\bibinfo{volume}{13}}, \bibinfo{pages}{14020},
  \doiprefix\url{10.1038/s41598-023-41320-0} (\bibinfo{year}{2023}).

\bibitem{Nakata2020}
\bibinfo{author}{Nakata, Y.}, \bibinfo{author}{Ito, Y.},
  \bibinfo{author}{Nakamura, Y.} \& \bibinfo{author}{Shindou, R.}
\newblock \bibinfo{journal}{\bibinfo{title}{Topological boundary modes from
  translational deformations}}.
\newblock {\emph{\JournalTitle{Phys. Rev. Lett.}}}
  \textbf{\bibinfo{volume}{124}}, \bibinfo{pages}{073901},
  \doiprefix\url{10.1103/PhysRevLett.124.073901} (\bibinfo{year}{2020}).

\bibitem{mieszczak2022}
\bibinfo{author}{Mieszczak, S.} \& \bibinfo{author}{Kłos, J.~W.}
\newblock \bibinfo{journal}{\bibinfo{title}{Interface modes in planar
  one-dimensional magnonic crystals}}.
\newblock {\emph{\JournalTitle{Scientific Reports}}}
  \textbf{\bibinfo{volume}{12}}, \bibinfo{pages}{11335},
  \doiprefix\url{10.1038/s41598-022-15328-x} (\bibinfo{year}{2022}).

\bibitem{berry_quantal_1984}
\bibinfo{author}{Berry, M.~V.}
\newblock \bibinfo{journal}{\bibinfo{title}{Quantal phase factors accompanying
  adiabatic changes}}.
\newblock {\emph{\JournalTitle{Proc. R. Soc. Lond.}}}
  \textbf{\bibinfo{volume}{392}}, \bibinfo{pages}{45--57},
  \doiprefix\url{10.1098/rspa.1984.0023} (\bibinfo{year}{1984}).

\bibitem{Kohn}
\bibinfo{author}{Kohn, W.}
\newblock \bibinfo{journal}{\bibinfo{title}{Analytic properties of {B}loch
  waves and {W}annier functions}}.
\newblock {\emph{\JournalTitle{Phys. Rev.}}} \textbf{\bibinfo{volume}{115}},
  \bibinfo{pages}{809--821}, \doiprefix\url{10.1103/PhysRev.115.809}
  (\bibinfo{year}{1959}).

\bibitem{Zak85}
\bibinfo{author}{Zak, J.}
\newblock \bibinfo{journal}{\bibinfo{title}{Symmetry criterion for surface
  states in solids}}.
\newblock {\emph{\JournalTitle{Phys. Rev. B}}} \textbf{\bibinfo{volume}{32}},
  \bibinfo{pages}{2218}, \doiprefix\url{10.1103/PhysRevB.32.2218}
  (\bibinfo{year}{1985}).

\bibitem{Collin01}
\bibinfo{author}{Collin, R.~E.}
\newblock \emph{\bibinfo{title}{Foundations for Microwave Engineering}}
  (\bibinfo{publisher}{Wiley-IEEE Press}, \bibinfo{year}{2001}).

\bibitem{Shockley39}
\bibinfo{author}{Shockley, W.}
\newblock \bibinfo{journal}{\bibinfo{title}{On the surface states associated
  with a periodic potential}}.
\newblock {\emph{\JournalTitle{Phys. Rev.}}} \textbf{\bibinfo{volume}{56}},
  \bibinfo{pages}{317--323}, \doiprefix\url{10.1103/PhysRev.56.317}
  (\bibinfo{year}{1939}).

\bibitem{Pozar11}
\bibinfo{author}{Pozar, D.~M.}
\newblock \emph{\bibinfo{title}{Microwave Engineering}}
  (\bibinfo{publisher}{John Wiley and Sons}, \bibinfo{year}{2011}).

\end{thebibliography}
\end{document}